\documentstyle[aps]{revtex}

\begin{document}
\title{Parallel Quantum Computation, the Library of Babel and Quantum Measurement
as the Efficient Librarian}
\author{G. Castagnoli}
\address{Elsag spa, Via G. Puccini 2, 16154 Genova, Italy}
\date{\today}
\maketitle

\begin{abstract}
The complementary roles played by parallel quantum computation and quantum
measurement in originating the quantum speed-up are illustrated through an
analogy with a famous metaphor by J.L. Borges.
\end{abstract}

\bigskip

\noindent Why there is a quantum speed-up is considered to be an interesting
open problem. This letter is an excerpt from the paper titled ``Performing
Quantum Measurement in Suitably Entangled States Originates the Quantum
Computation Speed Up'' [1]. That paper is rather lengthy, due to the need of
checking that all types of quantum algorithms found so far obey the speed-up
mechanism propounded. We shall summarize herebelow the justification of the
speed-up provided in [1].

Let us enter ``in media res''. One unnecessary but clarifying step of
Simon's algorithm is to measure the content of register $v$ (designated by $%
\left[ v\right] $ in the following) in the entangled state

\begin{equation}
\frac{1}{\sqrt{2^{n}}}\sum_{x=0}^{2^{n}-1}\left| x\right\rangle _{a}\left|
f\left( x\right) \right\rangle _{v}.
\end{equation}

\noindent State (1) is the result of reversibly computing the function $%
f\left( x\right) $ for all possible values of $x$, in quantum parallelism; $%
a $ and $v$ are two n-qubit registers containing the argument $x$ and the
respective function $f\left( x\right) $; $\left| x\right\rangle _{a}$
denotes an eigenstate of register $a$ (in the measurement basis), etc..
Given $B=\left\{ 0,1\right\} $, as well known $f\left( x\right) $ is a
function from $B^{n}$ to $B^{n}$ with the following properties:

\begin{itemize}
\item  \noindent for any $x$, there is one and only one $x^{^{\prime }}\neq x
$ such that $f\left( x\right) =f\left( x^{^{\prime }}\right) $;

\item  all such $x$ and $x^{^{\prime }}$ are evenly spaced by a constant $r$%
; namely, for all $x$, $\left| x-x^{^{\prime }}\right| =r$ (we are following
the simplified version of Simon's algorithm);

\item  given $x$, computing $f\left( x\right) $ requires poly(n) time,
whereas given a value $\stackrel{\_}{f}$ of $f\left( x\right) $, finding the
two values of $x$ such that $f\left( x\right) =f\left( x^{^{\prime }}\right)
=\stackrel{\_}{f}$ requires exp(n) time by classical computation -- the
function is hard to reverse.
\end{itemize}

The problem is to efficiently find $r$ by using a quantum computer that,
given $x$, computes $f\left( x\right) $ in poly(n) time. In fact, this
computer has already been used to reach state (1). Because of the character
of $f\left( x\right) $, measuring $\left[ v\right] $ in (1) yields an
outcome of the form

\begin{equation}
\frac{1}{\sqrt{2}}\left( \left| \stackrel{\_}{x}\right\rangle _{a}+\left| 
\stackrel{\_}{x}+r\right\rangle _{a}\right) \left| \stackrel{\_}{f}%
\right\rangle _{v},
\end{equation}

\noindent where $f\left( \stackrel{\_}{x}\right) =f\left( \stackrel{\_}{x}%
+r\right) =\stackrel{\_}{f}.$

Without entering into detail, the subsequent part of Simon's algorithm
consists in measuring $\left[ a\right] $ after performing the Hadamard
transform on the state of register $a$ in (2). The measurement outcome $z$
is such that $r\cdot z=0$; $r\cdot z$ denotes the module 2 inner product of
the two numbers in binary notation (seen as row matrices). By repeating the
overall process a poly(n) number of times, a number of constraints $r\cdot
z_{i}$ sufficient to identify $r$ is gathered with any desired probability
of success.

The role played by quantum measurement in originating the speed-up will be
discussed by using a special way of comparing quantum and classical
efficiency: the quantum computational cost of going from quantum state (1)
to quantum state (2) is benchmarked with the classical computational cost of
going through their (symbolic) descriptions. Such descriptions can be
visualized as the print-outs of (1) and (2), provided that $x,$ $f\left(
x\right) $, $\stackrel{\_}{x}$, etc. are substituted by proper numerical
values. This criterion is instrumental to achieving an a-posteriori self
evident result.

We shall instrumentally use the following way of thinking (opposite to our
view):

\begin{quote}
{\small quantum computation can produce a number of parallel outputs
exponential in register size, at the cost of producing one output, but this
``exponential wealth'' is easily spoiled by the fact that quantum
measurement reads only one output.}
\end{quote}

Let us examine the cost of classically deriving description (2) from
description (1). This latter can be visualized as the print-out of the sum
of 2$^{n}$ tensor products. Loosely speaking, two values of $x$ such that $%
f\left( x_{1}\right) =f\left( x_{2}\right) $, must be exp$\left( n\right) $
spaced. Otherwise such a pair of values could be found in poly$\left(
n\right) $ time by classical ``trial and error''.

The point is that the print-out would create a Babel Library\footnote{%
From the story ``The Library of Babel'' by J.L. Borges.} effect. Even for a
small $n$, it would fill the entire known universe with, say, $...$ $\left|
x_{1}\right\rangle _{a}\left| f\left( x_{1}\right) \right\rangle _{v}$ $...$
here, and $...$ $\left| x_{2}\right\rangle _{a}\left| f\left( x_{2}\right)
\right\rangle _{v}$ $...$ [such that $f\left( x_{1}\right) =f\left(
x_{2}\right) $] in Alpha Centauri. Finding such a pair of print-outs would
still require exp$\left( n\right) $ time. The capability of directly
accessing that ``exponential wealth'' would be frustrated by its
``exponential dilution''. This seems to be in match with the baffling
feeling inspired by Borges' story.

The quantum measurement of $\left[ v\right] $ instead, {\em distills} the
desired pair of arguments in a time \ linear in $n$, namely in the number of
qubits of register $v$.\footnote{%
It is a basic axiom of quantum measurement theory that the time required to
measure $\left[ v\right] $ is independent of state (1) entanglement --
entanglement is interaction free.} In fact, it does more than randomly
selecting one measurement outcome; by selecting {\em one} outcome, it
performs a logical operation (selecting the two values of $x$ associated
with the value of that outcome) crucial for solving the problem. The active
role played by quantum measurement in originating the speed-up,
complementary to the production of the parallel computation outputs, appears
to be self evident.

Ref. [1] also shows that performing or skipping $\left[ v\right] $
measurement in (1) is {\em equivalent}. It also formalizes the active role
played by quantum measurement. Given a suitably entangled state before
measurement, the constraint that there is a single measurement outcome
becomes a set of logical-mathematical constraints that represent the problem
to be solved (or the hard part thereof). Satisfaction of such constraints,
by the measurement outcome, amounts to having solved the problem. The
computational complexity of satisfying these constraints comes from
entanglement and is completely transparent to measurement time, which
justifies the speed-up.\footnote{%
Having found that the speed-up is an observable consequence of the
transition from the quantum to the classical world, naturally revamps the
quantum measurement problem. After having -- so to speak -- buried it, its
unexpected return might be coldly welcomed; while acknowledging this, we
should note that this time we are clearly dealing with a new and striking 
{\em fact}, not with a debatable interpretation of quantum measurement.}

Acknowledging the active role played by quantum measurement yields a more
realistic vision of what quantum computation is and is not.

It is not, as commonly believed, the quantum transposition of reversible
Turing-machine computation, where quantum measurement would only be needed
to {\em read} the output of a sequential computation process. In fact, we
have shown that quantum measurement plays a crucial role in efficiently {\em %
creating} that output. A quantum computation yielding a speed-up {\em is not}
a reversible computation, although reversibility is of course essential to
prepare the state before measurement.

Today, people is looking for non-sequential (e.g. topological) forms of
quantum computation. It is therefore a matter of some importance to
understand that the current ``quantum algorithms'' are already
non-sequential in character.

It is reasonable to think that detaching the notion of ``quantum algorithm''
from that of sequential computation -- a classical vestige -- is a
precondition for pursuing further developments at a fundamental level.

For example, let us consider the possibility of exploiting particle
statistics symmetrizations to achieve a quantum speed-up. Such
symmetrizations can be seen as projections on symmetrical (constrained)
Hilbert subspaces. There is no relation between a projection and sequential
reversible computation, namely a unitary evolution. If instead quantum
computation is (properly) seen as a projection on a constrained Hilbert
subspace, which amounts to solving a problem, then we have an analogy with
particle statistics symmetrizations to work with.

Thanks are due to the co-Authors of Ref. [1] for their consent to issue an
excerpt.

\noindent {\Huge References}

\begin{enumerate}
\item  G. Castagnoli, D. Monti, A. Sergienko, ``Performing Quantum
Measurement in Suitably Entangled States Originates the Quantum Computation
Speed Up'', arXiv: quant-ph/9908015 v2 14 Feb. 2000.
\end{enumerate}

\end{document}